\documentclass[floatfix,showpacs,prb,aps,twocolumn]{revtex4}

\usepackage{bm}
\usepackage{amsmath}
\usepackage{amssymb}
\usepackage{latexsym}
\usepackage{amsfonts}
\usepackage{epsfig}
\usepackage{psfrag}
\usepackage{color}

\newcommand{\ket}[1]{\left|#1\right>}
\newcommand{\bra}[1]{\left<#1\right|}

\newcommand{\expval}[1]{\left< #1 \right>}
\newcommand{\nn}{\nonumber\\}

\newcommand{\f}[1]{\mbox{\boldmath$#1$}}

\newcommand{\bea}{\begin{eqnarray}}
\newcommand{\ea}{\end{eqnarray}}
\newcommand{\eea}{\end{eqnarray}}

\newcommand{\abs}[1]{{\left| #1 \right|}}
\newcommand{\trace}[1]{{\rm Tr}\left\{ #1 \right\}}
\newcommand{\strace}[1]{{\rm\bf Tr}\left\{ #1 \right\}}

\newcommand{\binomial}[2]{\left(\begin{array}{c} #1\\ #2\end{array}\right)}

\definecolor{grey}{rgb}{0.5, 0.5, 0.5}
\definecolor{dgreen}{rgb}{0.0, 0.5, 0.0}
\definecolor{violet}{rgb}{0.5, 0.0, 0.5}
\definecolor{orange}{rgb}{1.0, 0.5, 0.0}

\begin{document}

\title{Counting Statistics in Multi-stable Systems}

\author{Gernot Schaller}
\email{gernot.schaller@tu-berlin.de}
\author{Gerold Kie{\ss}lich}
\email{gerold.kiesslich@tu-berlin.de}
\author{Tobias Brandes}

\affiliation{Institut f\"ur Theoretische Physik, Hardenbergstra{\ss}e 36,
Technische Universit\"at Berlin, D-10623 Berlin, Germany}

\begin{abstract}
Using a microscopic model for stochastic transport through a 
single quantum dot that is modified by the Coulomb interaction 
of environmental (weakly coupled) quantum dots, 
we derive generic properties of the full counting statistics for
multi-stable Markovian transport systems. We study the temporal crossover from
multi-modal to broad uni-modal distributions depending on the initial mixture, the long-term
asymptotics and the divergence of 
the cumulants in the limit of a large number of transport branches.  
Our findings demonstrate that the counting statistics of a single resonant level may be used to
probe background charge configurations.
\end{abstract}


\maketitle

\section{Introduction}

The coexistence of several stationary states for a
given set of parameters is typically referred to as the phenomenon of multi-stability. 
Multi-stable behavior is found in a wide variety of systems in different
disciplines of science, as e.g. 
biology \cite{ANG04}, chemistry \cite{JOH04},
neuroscience \cite{EAG01}, laser physics \cite{ARE82}, and semiconductor physics \cite{GAL08}.

In transport systems, multi-stability is characterized by the existence of more than
two distinct branches in the transport characteristics with hysteresis and switching
in between. 
Some prototype examples for
corresponding electronic systems are superlattices \cite{KAS94},
double-barrier resonant tunneling diodes \cite{MAR94}, and nano-electromechanical systems \cite{WIE08}.
If the transport is entirely governed by stochasticity, e.g. as in
single-electron transport \cite{BEE91a}, the current alone might not reveal the
multi-stable character and other more sensitive tools are required. 
As has been shown for bistable systems \cite{JOR04a}, the
counting statistics \cite{LEV93} may serve as such a tool.
Recently, in Ref.~\cite{FRI07} the measurement of a bi-modal
distribution of 
quantum dot tunneling has indicated the interplay of
fast and slow transport channels not visible in the current.

In this work, we present a generic approach to study Markovian 
transport systems with multi-stable behavior. 
Starting from a
microscopic model for one transport channel with an environmental
control system we derive a Master equation for Counting Statistics with
an arbitrary number of transport branches.
The resulting Liouville super-operator has a simple
and scalable block-tridiagonal structure. 
Even though there exists a unique steady state, the counting statistics and higher-order cumulants 
display clear signatures of multi-stability such as multi-modal
or broad distributions and diverging cumulants. 
We provide
results for the temporal evolution and long-term
asymptotics of the statistics and discuss the limit of
a large number of coexisting current branches analytically. 
We emphasize that our approach is not restricted to electrons as
transferred entities - in principle, stochastic multi-stable transport of any countable
object can be studied by this means.

\begin{figure}[b]
\includegraphics[width=0.45\textwidth,clip=true]{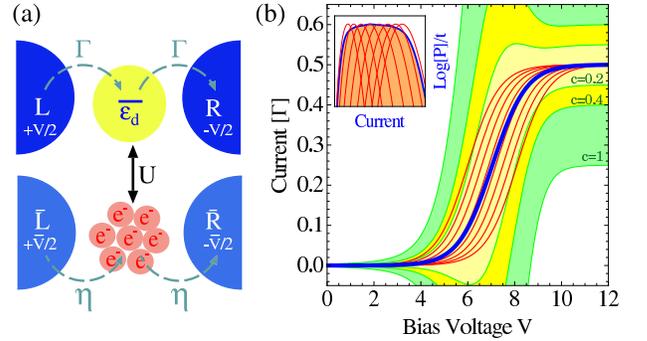}
\caption{\label{Fscheme}
(Color online) (a) Scheme of the model.
(b) Region of multi-stability: Current-voltage (Transport) characteristics for
$N=7$ control dots; thin red curves: $N+1$ individual currents $I_k$
($k=0, 1, \dots ,7$), thick
blue line: average current $\langle I\rangle$ which is actually observed; Borders of colored regions: standard
deviation $\pm c\sqrt{\left<\left< I^2 \right>\right>}/2$ of the current which becomes divergent for
$\eta\rightarrow 0$ in the multi-stability region. Inset: Broad uni-modal Distribution
function $\log{(P)}/t$ vs. current for large times and $V^*=$7 (thick
blue curve);
distributions for individual currents $I_k$ (thin red
curves). Parameters: $N=$7, $\epsilon_d$=3, $\beta$=3, $U$=0.15,
$\eta$=0.001. 
}
\end{figure}

\section{Illustrative picture}

A single transport channel (single resonant level, quantum dot) 
influenced by $k$ background charges distributed on a collection of $N$ sites
will experience an effective shift of its charged energy state 
($\epsilon_d \to \epsilon_d + k U$, where
$U$ is the Coulomb repulsion).
Attaching two reservoirs ($L$, $R$) held at different chemical potentials -- see Fig.~\ref{Fscheme}(a) --
will now induce transport through the dot at rate $\Gamma$, 
which strongly depends on the number of background charges.
Effectively, this leads to shifted currents in the transport channel
\bea\label{Ecurmulti}
I_k = \frac{\Gamma}{2} \bigg[ f_L(\epsilon_d+k U) - f_R(\epsilon_d+k U) \bigg]\,,
\eea
where $f_{L,R}(\omega)=\left[1+e^{+\beta(\omega\mp V/2)}\right]^{-1}$ 
denote the Fermi functions of the respective reservoir with bias voltage $V$, see Fig.~\ref{Fscheme}(b).
Coupling the background charges to different reservoirs $\bar L$, $\bar R$ with rate $\eta \ll \Gamma$ will
cause slow random switching  between the different transport channels $I_k$ (for 
two currents $k=0,1$ known as {\em random telegraph noise} \cite{KOG98}).
When the lead temperature is comparable to the Coulomb interaction $\beta^{-1} \approx U$, this gives
rise to a pronounced region of multi-stability around $V^* \approx 2 \epsilon_d + N U$ 
in the current-voltage characteristics, see Fig.~\ref{Fscheme}(b). 
In this region, the second-order cumulant of the current $\left<\left< I^2 \right>\right>$ 
diverges 
for $\eta\rightarrow 0$ as indicated. The inset shows the corresponding broad long-term 
distribution of currents at $V^*$ in comparison with the distributions of individual currents.


\section{Microscopic Model}

\subsection{Hamiltonian}

We consider the total Hamiltonian
\bea\label{Emodel}
H &=& \epsilon_d d^\dagger d + \sum_{i=1}^N \epsilon_i d_i^\dagger d_i 
+ \sum_{i=1}^N \sum_{j=i+1}^N U_{ij} d_i^\dagger d_i d_j^\dagger d_j\nn
&&+ \sum_{i=1}^N U_i d_i^\dagger d_i d^\dagger d + \sum_{ka} \omega_{ka} c_{ka}^\dagger c_{ka}\nn
&&+ \sum_{ka} t_{ka} \left[ d c_{ka}^\dagger + c_{ka} d^\dagger\right]\nn
&&+ \sum_{ka} \sum_{i=1}^N \tau_{ka}^i \left[ d_i c_{ka}^\dagger + c_{ka} d_i^\dagger \right]\,,
\eea
where $d$, $d_i$, and $c_{ka}$ annihilate electrons on the transport dot, the $i^{\rm th}$ control dot, and
the mode $k$ on lead $a$ (with energy $\omega_{ka}$), respectively.
In addition, we consider the symmetrized wide-band limit, where the transport dot
tunneling rates $\Gamma \equiv 2 \pi \sum_{k} \abs{t_{ka}}^2 \delta(\omega - \omega_{ka})$ and the 
control dot tunneling rates 
$\eta \equiv 2 \pi \sum_{k} \abs{\tau_{ka}^i}^2 \delta(\omega - \omega_{ka})$ 
become independent of energy and lead.
The parameters $U_i$ denote the Coulomb interaction between transport and control sites, whereas $U_{ij}$ 
represents repulsion between electrons within the control system.
We assume that the spectrum of the system Hamiltonian is only near but not exactly degenerate $U_i \approx U$, $U_{ij} \approx U_{\rm c}$, 
and $\epsilon_i \approx 0$.
These simplifications are not crucial for the occurrence of multi-stability, but rather allow for an analytic treatment in the following.

\subsection{Liouvillian}

We perform our analysis within the Born-Markov-secular approximation
scheme which can be alternatively \cite{SCH08,SCH09c} derived with a coarse graining
method in the limit of infinitely large coarse graining times $\tau$.
Provided the system energy spectrum is non-degenerate and the time
scales are larger than the inverse minimum level splitting, the Liouvillian couples only the 
diagonals of the density matrix in the system energy eigenbasis with
each other (see also \cite{BRE02}).
Since we are interested in observable effects of multi-stability in the current through the transport dot, 
we introduce a virtual detector in the right lead $R$
\cite{SCH09c} via the replacement in the tunneling Hamiltonian
$d c_{kR}^\dagger \to d c_{kR}^\dagger \otimes b^\dagger$ and
$c_{kR} d^\dagger \to c_{kR} d^\dagger \otimes b$,
where the detector operator $b^\dagger = \sum_n \ket{n+1} \bra{n}$ increases the 
detector outcome each time an electron is created in the right transport lead.
Treating the tensor product of dot and detector Hilbert spaces as the system, we arrive at an $n$-resolved master equation of the form
$\bra{n} \dot \rho \ket{n} \equiv \dot \rho^{(n)} = L_0 \rho^{(n)} + L_+ \rho^{(n-1)} + L_- \rho^{(n+1)}$, 
which couples different realizations of the dot density matrix -- each valid for different particle numbers $n$
measured in the detector -- with each other.
This coupled system can be further reduced by Fourier-transformation $\rho(\chi,t) \equiv \sum_n \rho^{(n)}(t) e^{i n \chi}$,
where $\chi$ is the counting field, which leads to
\mbox{$\dot \rho(\chi,t) = \left(L_0 + e^{+i \chi} L_+ + e^{-i \chi} L_-\right) \rho(\chi,t) \equiv L(\chi) \rho(\chi,t)$}.

Due to the permutational symmetry, it is convenient to denote the corresponding eigenstates for $N$ control sites
by $\ket{N,k,\ell;n_{\rm d}}$,
where $\ell \in \left\{1,2,\ldots, \binomial{N}{k}\right\}$ arbitrarily labels all the configurations
with $0 \le k \le N$ electrons distributed on the $N$ control sites, and $n_{\rm d}\in\{0,1\}$ denotes the 
occupation of the transport dot.
When we trace out the configuration of the control dots for a given total number of control charges $k$
by defining the $2\times 2$ matrix 
$\rho_{k}(\chi,t) \equiv \sum_{\ell} \bra{N,k,\ell} \rho(\chi,t) \ket{N,k,\ell}$,
the Liouvillian in this basis assumes for $0 \le k \le N$ the form
\bea\label{Etotal_liouville}
\dot \rho_{k}(\chi,t) &=& \Gamma {\cal L}^{\rm dot}(\chi,\epsilon_d+k U) \rho_{k}(\chi,t) + \eta {\cal L}_{0k}^{\rm control} \rho_{k}(\chi,t)\nn
&&+ \eta (N-k+1) {\cal J}_{k}^{\rm in} \rho_{k-1}(\chi,t)\nn
&&+ \eta (k+1) {\cal J}_{k}^{\rm out} \rho_{k+1}(\chi,t)\,,
\eea
where the newly introduced super-operators are $2\times 2$ matrices, which obey ${\cal J}_0^{\rm in} \equiv \f{0}$ and ${\cal J}_N^{\rm out} \equiv \f{0}$
at the boundaries.
This defines a $2(N+1)\times 2(N+1)$-dimensional Liouvillian super-operator 
$L(\chi)$ with a block-tridiagonal structure.
The detailed structure of the reduced Liouvillian super-operators follows from a rigorous microscopic derivation, it may however
also be understood from simple phenomenological reasoning:

{\bf (i)}
The multi-stable (fast) part has block-diagonal structure, where the $N+1$ block matrices correspond to 
the Liouvillian of a single resonant level -- shifted by the Coulomb interaction with $k$ control charges
\bea\label{Esinglecurmat}
{\cal L}^{\rm dot}(\chi,\omega) &\equiv& 
\left(
\begin{array}{cc}
- f_L(\omega) & f_L^-(\omega)\\
f_L(\omega) & - f_L^-(\omega)
\end{array}
\right)\nn
&&+\left(
\begin{array}{cc}
- f_R(\omega) & e^{+ i \chi} f_R^-(\omega)\\
e^{-i \chi} f_R(\omega) & - f_R^-(\omega)
\end{array}
\right)\,,
\eea
where $f_a^-(\omega) \equiv \left[1-f_a(\omega)\right]$.
Evidently, when $\eta=0$, these matrices give rise to the multi-stable currents in Eq.~(\ref{Ecurmulti}).
Since we have traced out the different control dot configurations, it also becomes obvious that the associated currents $I_k$ 
are actually $\binomial{N}{k}$-fold degenerate.
These degeneracies may be lifted (and thereby become observable) when the assumed symmetries of the Hamiltonian are absent.

{\bf (ii)}
The remaining part of the Liouvillian (which appears as slow when $\eta \ll \Gamma$) consists of 
the control system jump super-operators
\bea\label{Econtroljump}
{\cal J}_k^{\rm in} &\equiv&
\sum_{a \in \{\bar L, \bar R\}}
\left(\begin{array}{cc}
f_a((k-1) U_{\rm c}) & 0\\
0 & f_a\left((k-1) U_{\rm c}+U\right)
\end{array}\right)\,,\nn
{\cal J}_k^{\rm out} &\equiv& 
\sum_{a \in \{\bar L, \bar R\}}
\left(
\begin{array}{cc}
f_a^-(k U_{\rm c}) & 0\\
0 & f_a^-\left(k U_{\rm c}+U\right)
\end{array}
\right)\,,
\eea
which depend on the control system occupation number, 
and a trace-conserving part
${\cal L}_{0k}^{\rm control} \equiv - k {\cal J}_{k-1}^{\rm out} - (N-k) {\cal J}_{k+1}^{\rm in}$.
The scalar coefficients in Eq.~(\ref{Etotal_liouville}) arise, since for any control configuration
with $k-1$ charges there are \mbox{$N-k+1$} different possibilities to obtain a control configuration with $k$ 
charges.
Similarly, for a configuration with \mbox{$k+1$} charges, each single charge leaving the control sector constitutes
an equivalent jump channel~\cite{VOG10}.
In addition, we note that the control jump super-operators~(\ref{Econtroljump}) must assume diagonal form in the 
sequential tunneling regime for the basis chosen.


\subsection{Transport observables}

The probability for obtaining $n$ tunneled particles after time $t$ is
given by $P_n(t)=\trace{\rho^{(n)}(t)}$. It follows that the moments
of $P_n(t)$ may be directly obtained from the Fourier-transformed
Liouvillian by suitable differentiation of the moment generating
function (MGF)
\bea
M(\chi,t) = \strace{e^{L(\chi) t} \bar \rho}
\eea
(where $\strace{\left(\rho_0, \ldots, \rho_N\right)}\equiv\sum_{k=0}^N
\rho_k$)) with respect to the counting
field $\chi$.
The initial density matrix $\bar\rho$ is typically chosen as the
steady state $L(0)\bar\rho=0$, since one is usually interested in
long-term cumulants.
The matrix exponential is significantly harder to evaluate than the
matrix inverse, such that we consider the Laplace transform
\bea
\tilde M(\chi,z) = \strace{\frac{1}{z\f{1}-L(\chi)} \bar \rho}
\eea
of the MGF instead. 

For example, the moments of $P_n(t)$ are obtained via
\bea
\expval{n^k}(t)=(-i\partial_{\chi})^kM(\chi ,t).
\eea
The full distribution, however, is obtained by inverse Fourier transform
\bea
P_n(t) = \frac{1}{2\pi} \int\limits_{-\pi}^{+\pi} M(\chi,t)e^{-i n
\chi} d\chi .
\eea


\section{Results}

\subsection{Analytical steady state} 

\begin{figure}[t]
\includegraphics[width=0.45\textwidth,clip=true]{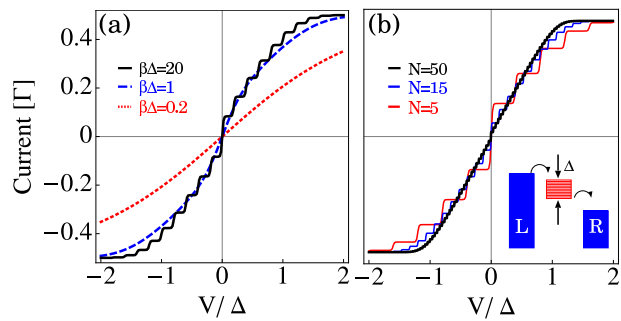}
\caption{\label{Fdistrib}
(Color online) Current-voltage characteristics for  a nontrivial non-thermal stationary state (\ref{Esteady}). 
(a) for various temperatures $\beta\Delta$ and $N=$10; at small temperatures ($\beta\Delta\gg $1) 
the number of control dots $N$ can be probed in a current measurement, whereas at $\beta\Delta\lesssim $1 one 
has to make use of counting statistics at $t\lesssim\eta^{-1}$ (see Fig.~\ref{Fprobdist}).
(b) for various $N$ and $\beta\Delta =$40; for $N\rightarrow\infty$, $U\rightarrow$0, $|V|<\Delta$ and $\beta\Delta\gg $1 the current becomes linear $I=\frac{e\Gamma}{2\Delta}V$ since transport takes 
place through a continuum with finite band-width $\Delta\equiv NU$ (as sketched in the inset).
}
\end{figure}

When transport and control dots are coupled to leads with the same chemical 
potential ($f_a=f_{\bar a}$) and $\epsilon_d=0$ as well as $U=U_{\rm c}$, the steady state of Eq.~(\ref{Etotal_liouville}) at $\chi=0$ is
$\bar\rho =C_N \left(\bar\rho_0,\dots\bar\rho_N\right)^T$,
where the partial vectors read
\bea\label{Esteady}
\bar\rho_k&=&
\binomial{N}{k} 
\left[\prod_{\ell=0}^{k-1} p_\ell\right]\left[\prod_{\ell=k+1}^{N} 2 - p_\ell\right]
\left(\begin{array}{c}
2-p_k\\
p_k
\end{array}\right)\nn
\eea
with $p_k\equiv f_L(k U)+f_R(k U)$ and $C_N$ follows from normalization.
The corresponding current $I = \trace{L'(0)\bar \rho}$ is the weighted sum of partial currents $I_k$ (\ref{Ecurmulti}). The current-voltage characteristic exhibits $2N$
steps for small temperatures $\beta\Delta\gg$1 ($\Delta\equiv NU$) which become smeared out for $\beta\Delta\lesssim $1 as shown in Fig.~\ref{Fdistrib}(a).
Therefore,  for sufficiently low temperatures we are able to probe the number $N$ just by a current measurement. 
At larger temperatures, however, this fails and
the counting statistics will provide a proper tool for that purpose (see Fig.~\ref{Fprobdist} and corresponding discussion). 
In the thermodynamic limit ($N\rightarrow\infty$, $U\rightarrow $0) 
such that the band-width $\Delta$ stays finite [spectrum becomes continuous as sketched in inset of Fig.~\ref{Fdistrib}(b)] the characteristic 
is linear for $\beta\Delta\gg$1 and $|V|<\Delta$ with Ohm's resistance of $2\Delta /\, e\Gamma$ [compare Fig.~\ref{Fdistrib}(b)].


\subsection{Full Counting Statistics}

The model (\ref{Emodel}) shows very rich behavior. We, however, choose
some limiting cases to illustrate the multistable properties in the following:

{\bf (i)} In the infinite bias limit $f_L\to 1$ and $f_R \to 0$, the MGF
coincides with that of a single resonant level, for which we obtain
\bea\label{Epdistsrl}
P_{n\ge0}(t) &=& \frac{e^{-\Gamma t}(\Gamma t)^{2n-1}}{2 (2n+1)!}\times\nn
&&\times\left[2n(2n+1)+(4n+2)\Gamma t + (\Gamma t)^2\right]\nn
\eea
and $P_{n<0}(t) = 0$, 
such that the counting statistics will not reveal any multi-stable properties
[compare Fig.~\ref{Fscheme}(b) for large $V$].

{\bf (ii)}
When the control leads are at infinite bias, i.e.
$f_{\bar L}(\omega) \to 1$ and $f_{\bar R}(\omega) \to 0$ such that \mbox{${\cal J}_k^{\rm in} = {\cal J}_k^{\rm out} = \f{1}$}, 
and the transport leads are at high bias ($f_R(\omega)\approx 0$), one may for sufficiently low temperatures have
$f_L(\epsilon_d+(k<\bar k)U)=1$ and $f_L(\epsilon_d+(k\ge\bar k)U)=0$ for some $\bar k \in \{1,\ldots,N\}$, 
which leads to only two different currents (bistable case).
The detailed form of the Liouvillian and its counting statistics then depends on $N$ and $\bar k$, but the whole class of bistable
models is amenable to analytic investigations.
In the simplest case of $\bar k = N=1$, we have for $\eta=0$ a bimodal distribution: Half of the distribution follows the
evolution of a single resonant level~(\ref{Epdistsrl}), and the other half remains localized at $n=0$ for all times.
The situation becomes non-trivial for finite $\eta$, which is reflected in the recursive relation for the 
Laplace transform 
$\tilde P_{n+1}(z) = {\cal F}(z) \tilde P_n(z)$ for $n\ge 2$, where ${\cal F}(z)$ has four different first order poles, 
such that -- unlike Eq.~(\ref{Epdistsrl}) -- the complexity of $P_n(t)$ will increase with $n$.

%

{\bf (iii)} 
Under the same (infinite and high bias, respectively) assumptions we may adjust bias voltage and temperature
such that we can distribute the left-associated Fermi-functions in an approximately equidistant manner between zero and one
(such as e.g. $f_L(\epsilon_d)=1$, $f_L(\epsilon_d+U)=2/3$, $f_L(\epsilon_d+2U)=1/3$, and $f_L(\epsilon_d+3U)=0$ for $N=3$),
we can analytically extract the current $\left<\left< \dot n^1(t) \right>\right> = \frac{\Gamma}{4}$ and
the long-term scaling of the next higher cumulants (for $N\neq 0$) of $P_n(t)$
\bea
\left<\left< n^2(t) \right>\right> &\to&  \frac{\Gamma \left(3 N \eta^2 + 3 N \eta \Gamma + \Gamma^2\right)}{16 N \eta \left(\Gamma+\eta\right)}t\,,\nn
\left<\left< n^3(t) \right>\right> &\to & \frac{\Gamma}{64} \left[7 + \frac{3 \Gamma^2\left(2 \Gamma + 3 \eta\right)}{\eta N \left(\Gamma + \eta\right)^2}\right] t\,.
\eea
These expressions demonstrate that the higher cumulants diverge for small $\eta$ in the long time limit.
In the limit of an infinite number of multi-stable channels ($N\to\infty$) however, this divergence is overshadowed 
by the exponentially large degeneracy of intermediate currents: 
If the control jump matrices in Eq.~(\ref{Etotal_liouville}) did not scale with $N$, the divergence of all higher
than second cumulants would persist also in the limit $N\to\infty$.

\begin{figure}[t]
\includegraphics[width=0.45\textwidth,clip=true]{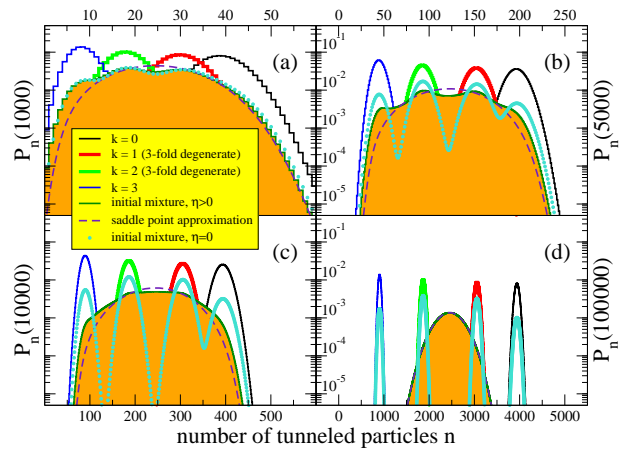}
\caption{\label{Fprobdist}
(Color online) Probability distribution $P_n(t)$ for the number of
tunneled particles $n$ at different times $t$ (orange region).
For increasing times [(a) $\rightarrow$ (d)] the average of the distribution moves
linearly towards larger $n$.
It proceeds a crossover from multi-modal to unimodal with a transition
time of $\eta^{-1}$ and becomes (nearly) a broad Gaussian for $t\gg\eta^{-1}$ (d).
For $\eta =0$ even the long-term distribution depends on the initial
mixture (symbols) here chosen as analytic continuation of $\bar\rho$
to $\eta =0$, which reflects in the different peak weights.
For comparison: Distributions $P_n^{(k)}(t)$ for individual currents $I_k$ (blue, bold green, bold red, and
black curves in the background).
The saddle-point approximation (dashed lines) only captures the
long-term behavior $t\gg\eta^{-1}$.
Parameters: $N=3$, $\epsilon_d=1$, $\Gamma=0.1$, $\eta=0.0001$, $U=1$,
$\beta =1$, $V=V^*=5$, $\bar V \to \infty$.
}
\end{figure}

{\bf (iv)} 
Without these assumptions, we can still numerically perform both the
inverse Laplace transform of $\tilde M(\chi,z)$ and a Fourier integral
to obtain $P_n(t)$, which is typically evaluated using the saddle-point approximation
\cite{SCH09c} [compare also inset of Fig.~\ref{Fscheme}(b)].
The result in the multi-stable bias regime of interest is shown for
different times in Fig.~\ref{Fprobdist}.
Choosing $\bar\rho$ for $\eta >0$ as initial mixture, the statistics
is multi-modal with $N+1$ maxima for $t \lesssim \eta^{-1}$
[Figs.~\ref{Fprobdist}(a) and (b)]. 
In contrast, the statistics becomes
uni-modal for $t \gtrsim \eta^{-1}$ [Figs.~\ref{Fprobdist}(c) and (d)].
In the long-term limit, the distribution for $\eta >0$ evolves essentially into a
broad Gaussian (d). 
This is a general property of systems with cumulants
linearly evolving in time.
In contrast, when $\eta =0$, $P_n(t)$ will depend on the initial
configuration for all times. For example, when one initializes in one
of the subspaces $k=0,\dots, N$, one will observe the distributions
$P_n^{(k)}(t)$ for the individual currents $I_k$ (curves in the
background in Fig.~\ref{Fprobdist}).
Starting in a statistical mixture for $\eta =0$ yields a multi-modal distribution
even in the long-term limit (symbols in Fig.~\ref{Fprobdist}) and leads to a divergence of all
higher-order cumulants. 


\subsection{Experimental parameters} 

For the observation of a multi-modal distribution (e.g. bi-modal 
in 
Ref.~\cite{FRI07}) the measurement time must lie between $\Gamma^{-1}$ and $\eta^{-1}$.
The rates in Ref.~\cite{FRI07} are of the order of $\Gamma \approx 1 \rm kHz$ and $\eta \approx 1 \rm Hz$, respectively, 
such that the time of measurement can be estimated between $1$ ms and $1$ s. 
For a distance of a hundred $\rm nm$ between transport and control system, the Coulomb interaction strength 
in GaAs can be estimated to $U \approx 1.2 \rm meV$.
Provided the picture of a single transport level is still valid (i.e., for a significantly larger on-site Coulomb interaction energy),
pronounced multi-stability should be observable around temperatures of $T \approx 14 \rm K$.
Larger distances or screened Coulomb interactions would lead to lower temperatures.


\section{Conclusions and Outlook} 

We have studied stochastic multi-stable transport
in terms of an $n$-resolved Master equation with a simple and scalable block-tridiagonal Liouville super-operator 
(\ref{Etotal_liouville}).
Multi-stability can
be revealed by the full counting statistics even when the first moments
are insensitive: For measurement times smaller than the switching rate
between the distinct transport channels the distributions are
multi-modal when the initial state is a mixture of the multiple steady
states for $\eta =0$ (this is the case for $\bar\rho$ for finite $\eta$).
This enables direct access to the number of decoupled
non-degenerate subspaces. 
For longer times or degenerate subspaces
this is not possible. 
However, unusually large
higher-order cumulants may point towards intrinsic multi-stability.
In case of sufficiently low temperatures multi-stable distributions
may become effectively bi-stable.

However, if the initial mixture is strongly localized within one of
the multiple subspaces (this would be a consequence of a
projective measurement), a particle-number measurement would result in
the associated
current with high probability and all other currents with low
probabilities.
Consequently, a sequence of repeated measurements would yield the
switching dynamics observed for single-charge counting detectors \cite{GUS06,FRI07,FLI09}.

We finally remark that multi-stable behavior can also emerge due to the effect of coherences~\cite{SCH09c}.


\acknowledgments

We thank C. Emary and W. Belzig for useful discussions. 
Financial support by the DFG (project \mbox{BR 1528/5-1}) is gratefully acknowledged.


\bibliographystyle{/usr/share/texmf-dist/bibtex/bst/revtex/apsrev}
\bibliography{/home/kieslich/references}

\end{document}